\documentclass[nofootinbib,preprintnumbers,prd,superscriptaddress,twocolumn]{revtex4}

\usepackage[titletoc]{appendix}
\usepackage{color}
\usepackage{hyperref}

\usepackage[rightcaption]{sidecap}
\usepackage{subfigure}
\usepackage{comment}
\usepackage[inkscapelatex=false]{svg}
\usepackage{graphicx}
\usepackage{transparent}
\usepackage{amsmath}
\usepackage{amsfonts}
\usepackage{amsthm}
\usepackage{mathrsfs}
\usepackage{amssymb}
\usepackage{epsfig}
\usepackage{amscd}
\usepackage{dcolumn}
\usepackage{bm}
\usepackage{natbib}
\usepackage{url}
\usepackage{xspace}

\usepackage{svg}

\usepackage{dcolumn}

\usepackage{array}
\usepackage{ctable}
\usepackage{multirow}
\usepackage{siunitx}
\usepackage{longtable}
\usepackage{tabularx}
\usepackage{booktabs}

\usepackage{amsthm}
\usepackage{mathrsfs}

\usepackage{epsfig}
\usepackage{amscd}

\usepackage{bm}
\usepackage{natbib}
\usepackage{url}
\usepackage{xspace}
\usepackage{kantlipsum}

\graphicspath{{Graphics/}}

\def\be{\begin{equation}}
\def\ee{\end{equation}}
\def\bea{\begin{eqnarray}}
\def\eea{\end{eqnarray}}

\definecolor{vividviolet}{rgb}{0.62, 0.0, 1.0}
\definecolor{amaranth}{rgb}{0.9, 0.17, 0.31}
\definecolor{palatinateblue}{rgb}{0.15, 0.23, 0.89}
\definecolor{brightpink}{rgb}{1.0, 0.0, 0.5}
\definecolor{cornflowerblue}{rgb}{0.39, 0.58, 0.93}
\definecolor{deepcarminepink}{rgb}{0.94, 0.19, 0.22}
\definecolor{radicalred}{rgb}{1.0, 0.21, 0.37}

\hypersetup{ linktoc=all,
    colorlinks, linkcolor={palatinateblue},
    citecolor={brightpink}, urlcolor={amaranth}
}

\date{\today}

\newcommand\fverb{\setbox\pippobox=\hbox\bgroup\verb}
\newcommand\fverbit{\egroup\item[\fbox{\unhbox\pippobox}]}

\newbox\pippobox

\begin{document}

\title{Constructing regular black holes from multi-polytropic equations of state}

\author{Seyed Naseh Sajadi}
\email{naseh.sajadi@gmail.com}
\affiliation{Strong Gravity Group, Department of Physics, Faculty of Science, Silpakorn University, Nakhon Pathom 73000, Thailand.}

\author{Supakchai Ponglertsakul}
\email{supakchai.p@gmail.com}
\affiliation{Strong Gravity Group, Department of Physics, Faculty of Science, Silpakorn University, Nakhon Pathom 73000, Thailand.}

\author{Orlando Luongo}
\email{orlando.luongo@unicam.it}
\affiliation{University of Camerino, Via Madonna delle Carceri, Camerino, 62032, Italy.}
\affiliation{SUNY Polytechnic Institute, 13502 Utica, New York, USA.}
\affiliation{INAF - Osservatorio Astronomico di Brera, Milano, Italy.}
\affiliation{Istituto Nazionale di Fisica Nucleare (INFN), Sezione di Perugia, Perugia, 06123, Italy.}
\affiliation{Al-Farabi Kazakh National University, Al-Farabi av. 71, 050040 Almaty, Kazakhstan.}

\begin{abstract}
Regular black holes are imagined as solutions to Einstein's field equations, with no singularities, albeit characterized by the presence of an internal structure. With the intention not to use non-linear electrodynamics, we here propose to obtain new classes of solutions that can also satisfy the Tolman-Oppenheimer-Volkoff (TOV) equations, plus adding a non-zero core. Thus, we present regular black holes as solutions to the TOV equations using multipolytropic equations of state and investigate whether these solutions behave, tuning the underlying free parameters. Our analysis demonstrates that, within specific parameter ranges,  repulsive gravity effects may occur in precise regions. Accordingly, black hole remnants are also investigated, showing that, under certain circumstances, they may turn into dark energy sources in view of the corresponding repulsive gravity effects, located outside the horizons. Moreover, quite remarkably,  critical sets of parameters imply that solutions may exhibit transitions to regular repulsive relativistic compact objects from black hole behaviors. Finally, we explore the interpretation of these regular black hole solutions in terms of topological thermodynamic defects.
\end{abstract}

\maketitle

\section{Introduction}

Even though general relativity (GR) has successfully passed numerous experimental tests, theoretical inconsistencies seem to jeopardize Einstein's field equations at both infrared and ultraviolet regimes. For example, in the presence of strong gravitational fields, the lack of a complete quantum gravity theory \cite{Kiefer:2023bld, Kieferbook} and the issue of singularities are clearly unresolved challenges. Similarly, at cosmological scales, open questions remain toward cosmological tensions  \cite{Freedman:2021ahq} and the nature of the universe's dark components \cite{Copeland:2006wr, Oks:2021hef,my1,my2}. 

Accordingly, the Penrose theorems state that gravitational collapse inevitably leads to curvature singularities \cite{Penrose:1964wq} which GR fails to be predictive. Nevertheless, singularities are generally associated with phase transition and not common in nature with a notable exception of the Big Bang \cite{Penrose:1969pc}. Consequently, the \emph{singularity problem} is closely tied to the existence of black holes (BHs), whose reality has been strongly supported by the direct detection of gravitational waves and BH shadows \cite{LIGOScientific:2016aoc, EventHorizonTelescope:2019dse}.

In view of a quantum gravity theory and of the fact that singularities are not physical frameworks, fully understanding BH properties may be required to avoid singularities or modify the corresponding classical theory, including quantum effects\footnote{For effective quantum BH pictures, see \cite{Zhang:2023okw}. }. Remarkably, singularities can be avoided if the energy conditions, assumed by the singularity theorems are violated at least in some spacetime regions. Regular regions can replace singularities with some exotic form of matter that violates the strong energy condition. Following the pioneering contribution by Sakharov and Gliner  \cite{sakharov,gliner}, the first idea suggesting that singularities could be avoided is based on including a de Sitter core\footnote{More precisely, de Sitter cores may be associated with vacuum energy contributions that avoid the presence of singularities. For a generalization of de Sitter solutions see \cite{Dymnikova:2004zc}.}. Following this recipe, a wide class of regular BH (RBH) solutions has been released in the literature, see e.g. \cite{Lan:2023cvz, Torres:2022twv,afterT1,afterT2}.

In this respect, the first smooth RBH, based on this prescription, is first proposed by Bardeen \cite{bardeen}. For completeness, RBHs have natural horizons but no singularities, placed at the center, i.e., the solution tends to be a de Sitter core solution and, generally, to a Schwarzschild solution asymptotically. In addition, Ayon-Beato and Garca successfully interpreted the Bardeen BH in the framework of non-linear electrodynamics \cite{Ayon-Beato:1998hmi}. Later, many other RBHs have been obtained invoking the aforementioned nonlinear electrodynamics in GR \cite{Dymnikova:1992ux, Novello:2000km, Bronnikov:2000yz, Hayward:2005gi, Sajadi:2017glu} and modified gravities, i.e., 4D Einstein Gauss-Bonnet gravity \cite{Kumar:2020uyz}, $f(R)$ gravity \cite{Rodrigues:2015ayd} and $f(R,T)$ gravity \cite{Tangphati:2023xnw}.

Significant progress has been made in the study of regular BH physics over the past decades. In particular, RBHs have been explored in the context of quantum gravity, with possible quantum origins linked to modifications of the uncertainty principle \cite{Wilson}. Additionally, RBHs have been derived from pure gravitational theories in higher-dimensional frameworks \cite{Bueno:2024dgm}. Various aspects of their properties have been investigated, including thermodynamics \cite{Ma:2014qma,Balart:2017dzt}, inner-horizon instability \cite{Carballo-Rubio:2022kad, Bonanno:2020fgp}, and observational signatures such as their shadows \cite{Li:2013jra,Dymnikova:2019vuz,Ling:2022vrv,Hendi:2020knv}. Furthermore, studies have examined RBH quasinormal modes \cite{Flachi:2012nv,Cai:2020kue,Sajadi:2019hzo}, superradiance effects \cite{Yang:2022uze}, and the potential existence of primordial RBHs \cite{Calza:2024fzo}.

Quite remarkably, it has been shown that the matter content of an RBH can exhibit a polytropic equation of state \cite{Shojai:2022pdq}. Almost all polytropic models assume an isotropic pressure, i.e., the tangential and radial components are the same. {Recently, RBH is obtained by considering Kiselev black hole \cite{Kiselev:2002dx} surrounding by anisotropic fluid \cite{Santos:2024vby}.}

In general, however, all principal pressures can have double-polytropic equations of state, so, several authors have investigated anisotropic models that provide different pressures for the tangential and radial components, see e.g. \cite{Chaisi:2006rg,Dev:2000gt,Dev:2003qd,Ivanov:2002jy}. For example, although the radial pressure vanishes at the surface of a compact star, one could postulate a tangential pressure to exist. While the latter does not alter the spherical symmetry, it may create some streaming fluid motions \cite{Harko:2002pxr,Mak:2001eb}. In this respect, the Reissner-Nordstr\"om solution is another relativistic solution that is based on the existence of anisotropic stress tensor. A charged BH is associated with negative radial pressure, while the tangential pressure remains positive \cite{Ortin:2015hya}. This anisotropic pressure originates from the fact that the particle's phase space distribution function depends on coordinates and velocities only through the constants of motion (energy and angular momentum). If the distribution function depends on the magnitude of the angular momentum, this will lead to an anisotropic pressure, even if the system is spherically symmetric.

Motivated by the fact that there are two different ways to construct RBH, i.e., by solving the field equations associated with a special source of matter, whereas the second is to write a desired RBH solution at first, then determining \emph{a posteriori} the corresponding action of matters. Here, we obtain RBH solutions inspired by solving the relativistic version of the classical hydrostatic equilibrium equation inspired by the TOV spacetime \cite{tolman,openhim} with a multi-component polytropic equations of state, hereafter multi-polytropic equations of state.

Hence, we allow non-zero anisotropic contributions to the pressure by assuming different tangential and radial components in the energy-momentum tensor and consider different equations of state, emphasizing specific cases that can be of particular interest. Here, the important point is that the regularity of a BH comes from the tangential pressure of its constituent matter, especially from the nonlinear part of the tangential pressure. This means that the multi-polytropic anisotropic perfect fluid is a suitable source for RBHs.

Traditionally, the polytropic equation of state has been used to describe a
completely degenerate gas in white dwarfs and a completely convective star. Moreover, different parts of astrophysical objects may obey polytropic behavior but with different polytropic indices. It is therefore reasonable to assume that an equation of state which is a mixture of different polytropes (i.e. a multi-polytropic equation of state) can be a better fit to the real equation of state. A multi-polytropic equation of state that is softer at low densities and stiffer at high densities may be appropriate for describing a hybrid astrophysical object. Afterwards, a more general equation of state provides a better fit, thereby increasing the freedom to describe naturally occurring astronomical objects.

The paper is organized as follows. In Section \ref{sec2}, we solve the generalized TOV equation with an anisotropic equation of state.
In sections \ref{subsec2.1} and \ref{subsec2.3}, we present two RBHs and study the physical properties of the solutions, and the last section \ref{concluding} reflects the concluding remarks of our work.

\section{The anisotropic TOV equation}\label{sec2}

We here start considering statically and  spherically symmetric spacetime with coordinates $(t,r,\theta,\phi)$, describing gravitating relativistic object 
\begin{equation}\label{metricasl}
ds^{2}=-f(r)dt^2+h(r)dr^2+r^2d\theta^2 +r^2\sin^{2}\theta d\phi^2 ,
\end{equation}
where $f(r)$ and $h(r)$ denote the  unknown lapse and shift functions, respectively. From Einstein's field equations, $
\mathcal{R}_{\mu\nu}-\frac{1}{2}g_{\mu\nu}\mathcal{R}=8\pi G\mathcal{T}_{\mu\nu}$, considering an anisotropic fluid, we write the energy-momentum tensor,
\begin{equation}
\mathcal{T}^{\mu}_{\nu}=\text{diag}(-\rho,P_{r},P_{t},P_{t}),
\end{equation}
here $\rho$, $P_{r}$, and $P_{t}$ are the energy density, radial and transverse components of the pressure, respectively. 

Let's introduce the mass variable $M(r)$ via
\begin{align}
    h(r) &= \left(1-\frac{2GM(r)}{r}\right)^{-1},
\end{align}
where $M(r)$ is interpreted as the gravitational mass inside radius $r$, satisfying 
\begin{equation}
M(r)=\int_{0}^{r}4\pi \tilde{r}^ 2\rho(\tilde{r})d\tilde{r},
\end{equation}
that also relates the density, $\rho$, and the mass, $M$.

From the TOV spacetime and adopting $\nabla_{\mu}\mathcal{T}^{\mu}_{r}=0$, we infer the equation for the radial and tangential pressures,
 \begin{equation}\label{GTOVE}
 -P^{\prime}_{r}=\dfrac{G(\rho+P_{r})(M+4\pi r^{3}P_{r})}{r(r-2M)}+\dfrac{2(P_{r}-P_{t})}{r},
 \end{equation}
where prime denotes derivative with respect to $r$. Remarkably, the effect of anisotropy is evident within the last term of \eqref{GTOVE}. To seek an RBH class of solution, we first assume the following barotropic equations of state (EoS) \cite{Sajadi:2016hko, Riazi:2015hga, Sajadi:2023ybm}
\begin{align}
&P_{r}= \omega \rho +\bar{\omega}\dfrac{\rho^{n}}{\rho_{0}^{n-1}},\label{eos1}\\
&P_{t}=\omega_{1}\rho+\omega_{2}\dfrac{\rho^{m}}{\rho_{0}^{m-1}},\label{eos2}
\end{align}
here $\rho_{0}>0$ is the central density, $\omega, \bar{\omega}, \omega_{1}$ and $\omega_{2}$, are the dimensionless EoS parameters. 

The first term, appearing in \eqref{eos1}--\eqref{eos2}, is the widely-used linear EoS, where the sound speed contribution is constant, whereas the second term allows for a polytropic behavior, adopted to characterize stars and/or interiors, e.g. leading to the Lane-Emden equation in the theory of stellar structure. In addition, the polytropic EoS is also used in several cosmological realms, see e.g. \cite{clayton}.

To obtain the solutions with $\mathcal{T}^{t}_{t}=\mathcal{T}^{r}_{r}$ resulting into the Schwarzschild coordinates, namely $f=h^{-1}$, we require $\omega=-1,\bar{\omega}=0$. Thus, the anisotropic fluid consists of the following components of the pressure:
\begin{equation}
P_{r}=-\rho,\;\;\;\;\;P_{t}=\omega_{1}\rho+\omega_{2}\dfrac{\rho^{m}}{\rho_{0}^{m-1}}.
\end{equation}
 
Plugging the above assumptions into \eqref{GTOVE}, one obtains $\rho$,
 \begin{equation}\label{energydens}
 \rho(r)=\dfrac{\rho_{0}}{\left[c_{1}(\rho_{0}r^{2(\omega_{1}+1)})^{m-1}-\dfrac{\omega_{2}}{\omega_{1}+1}\right]^{\frac{1}{m-1}}},
 \end{equation}
where $c_{1}$ is an integration constant. Near the origin ($r\to 0$), the energy density reaches its maximum and reads 
\begin{equation}
\rho(r\to 0)\sim \dfrac{\rho_{0}}{\left(\dfrac{-\omega_{2}}{\omega_{1}+1}\right)^{\frac{1}{m-1}}},
\end{equation}
{Since $\rho \to \rho_0$ as $r\to 0$, therefore, we require that $\omega_{2}=-\omega_{1}-1$.}

At large $r$, the energy density behaves as
\begin{equation}
\rho(r\to \infty)\sim \dfrac{1}{c_{1}^{{\frac{1}{m-1}}}r^{2(\omega_{1}+1)}}.
\end{equation}

Now if we assume $m>0,c_{1}>0$ and $\omega_{1}>0$, the energy density is always positive. 

The tangential pressure vanishes in
\begin{equation}
P_{t}=0,\;\;\;\;\;\to\;\;\;\; r=\left(\dfrac{\rho_{0}^{-m+1}}{\omega_{1}c_{1}}\right)^{\frac{1}{2(\omega_{1}+1)(m-1)}}.
\end{equation} 
At this stage, by inserting \eqref{energydens} into the Einstein field equations, one can solve and determine the metric functions. In the following sections, we intend to set precise values for the free parameters, $m$ and $\omega_{1}$, to present singularity-free solutions that appear physically, i.e., matchable with compact objects.
  
\section{The first solution}\label{subsec2.1}

\begin{figure*}
\centering
\includegraphics[width=0.95\columnwidth]{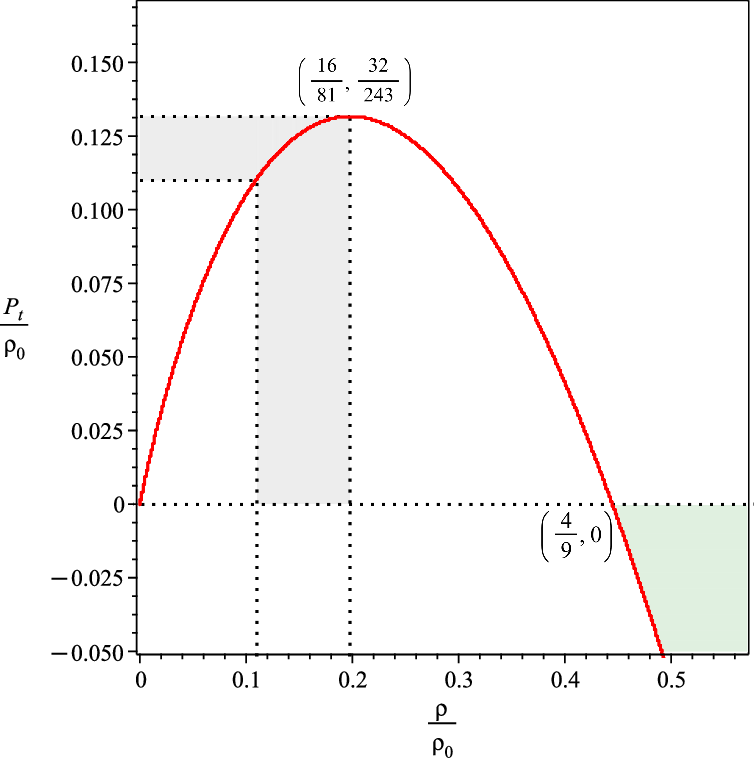}
\includegraphics[width=0.95\columnwidth]{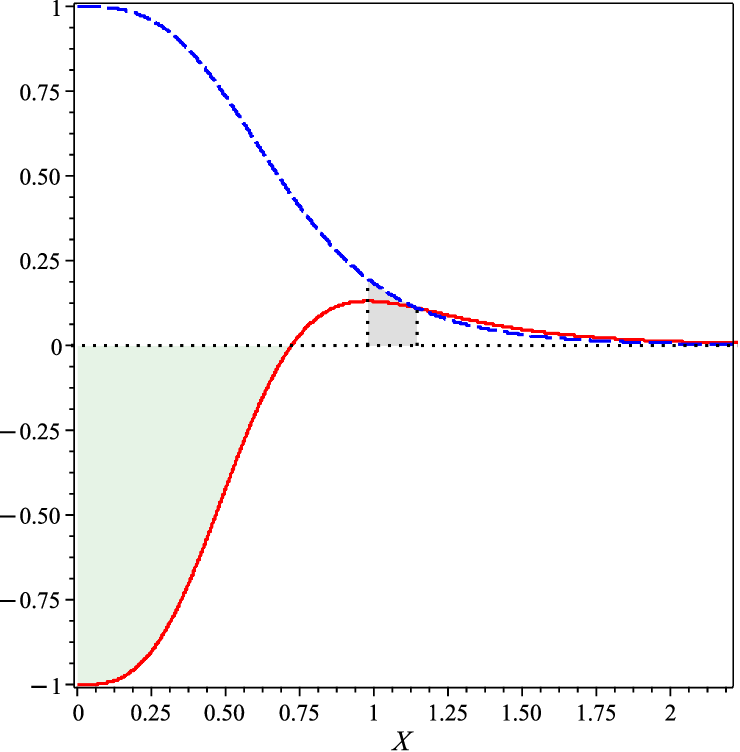}
\caption{Left: The behavior of $P_{t}/\rho_{0}$ in terms of $\rho/\rho_{0}$ for $m=3/2$ and $\omega_{1}=2$. Right: The behavior of $P_{t}/\rho_{0}$ (solid line) and $\rho/\rho_{0}$ (dash line) in terms of $X=(\pi\rho_{0}M^2)^{\frac{1}{3}}r/M$. The gray shaded regions are the allowed regions for which $1/9\leq\rho/\rho_{0}\leq 16/81$ (left) and $(15/8)^{1/3}\leq X \leq (3/2)^{1/3}$(right). In the green-shaded regions, SEC violated.} 
\label{wpert0}
\end{figure*}

An immediate and very relevant case is offered by the choice of the free constant made as:
\begin{subequations}
    \begin{align}
&m=\frac32,\\
&\omega_{1}=2.        
    \end{align}
\end{subequations}
Here, we notice that any value given to $m$ and $\omega_{1}$ leads to a possible solution. However, not all solutions yield a finite value for BH mass. In our choices of constants, we present those values that provide a finite BH mass.
Immediately, the tangential pressure,  $P_{t}$, is given as follows
\begin{equation}\label{regular0}
P_{t}=2\rho -\dfrac{3\rho^{\frac{3}{2}}}{\sqrt{\rho_{0}}} >0,\;\;\to\;\;\begin{cases}
0<\rho<\dfrac{4}{9}\rho_{0},\\
\\
\left(\dfrac{1}{2c_{1}\sqrt{\rho_{0}}}\right)^{\frac{1}{3}}<r<\infty.
\end{cases}
\end{equation}
The behavior of $P_{t}/\rho_{0}$ in terms of $\rho/\rho_{0}$ is shown in Fig. \eqref{wpert0}. For $\rho={4\over9}\rho_{0}$, the tangential pressure vanishes beyond which the strong energy condition (SEC) is violated and the gravitational attraction becomes gravitational repulsion, i.e. for the region near the center $0\leq r\leq \left(\frac{1}{2c_{1}\sqrt{\rho_{0}}}\right)^{\frac{1}{3}}$. For ${1\over9}\rho_{0}<\rho<{4\over9}\rho_{0}$, the dominant energy condition (DEC) is satisfied, i.e., $\rho>P_{t}$. For $\rho={16\over81}\rho_{0}$ or $r=({5\over4}c_{1}\sqrt{\rho_{0}})^{\frac{1}{3}}$, the tangential pressure has a maximum value at $P_{t,max}={32\over243}\rho_{0}$. For ${4\over81}\rho_{0}<\rho<{16\over81}\rho_{0}$ the sound speed is between $0\leq v_{s}=\sqrt{dP/d\rho}\leq 1$. Therefore, the allowed region (shaded region in gray) for which the above conditions are satisfied is: ${1\over9}\rho_{0}<\rho<{16\over81}\rho_{0}$, corresponding to $(2/c_{1}\sqrt{\rho_{0}})^{\frac{1}{3}}\leq r\leq (5/4c_{1}\sqrt{\rho_{0}})^{\frac{1}{3}}$. We also display $P_t/\rho_0$ and $\rho/\rho_0$ as function of $X=(\pi\rho_{0}M^2)^{\frac{1}{3}}r/M$ in the right panel of Fig.~\eqref{wpert0}. We observe that the energy density is regular at the center as expected. Then, it becomes decreasing as $r$ increases. On the other hand, the tangential pressure increases (from negative) with $r$ and vanishes at $X=0.7211.$ This agrees with the condition discussed in \eqref{regular}. The pressure reaches its maximum at $X=0.9787.$

Furthermore, the metric function is obtained as follows
\begin{equation}\label{hywardlike}
f(r)=1-\dfrac{8\pi \sqrt{\rho_{0}}}{3c_{1}r}+\dfrac{8\pi\sqrt{\rho_{0}}}{3c_{1}r(1+c_{1}\sqrt{\rho_{0}}r^{3})}.
\end{equation}
To find the physical significance of the parameter $c_{1}$, we consider the behavior of the solution at $r\to\infty$ as follows 
\begin{equation}
f(r)\sim 1-\dfrac{8\pi\sqrt{\rho_{0}}}{3c_{1}r}+\dfrac{8\pi}{3c_{1}^{2}r^{4}}+\mathcal{O}(r^{-7})
\end{equation}
It is easy to figure out that $c_{1}$ corresponds to a parameter with dimension $[c_{1}]=L^{-2}$. 

This solution describes the gravitational field of a BH if an outer horizon exists outside the radius of the compact object \cite{Liu:2021fit}. To obtain the radius of the object, we focus on its gravitational mass, so having as the total mass of the relativistic gravitating system, 
\begin{equation}\label{eqmassHyward}
M=4\pi\int_{0}^{\infty}r^{2}\rho(r)dr=\dfrac{4\pi\sqrt{\rho_{0}}}{3c_{1}},
\end{equation}
and assuming that the above mass is equal to $M=\rho_{0}R^{3}$, then we obtain
\begin{equation}
R\sim\left(\dfrac{M}{\rho_{0}}\right)^{\frac{1}{3}}\sim\left(\dfrac{4\pi}{3c_{1}\sqrt{\rho_{0}}}\right)^{\frac{1}{3}}.
\end{equation}
This can be interpreted as an effective radius if a large mass approximation is assumed.

Nevertheless, the roots of the metric function \eqref{hywardlike} are 
\begin{align}
r_{0}=&\dfrac{\mathcal{A}^{\frac{1}{3}}}{18c_{1}}+\dfrac{128\pi^2\rho_{0}}{9c_{1}\mathcal{A}^{\frac{1}{3}}}+\dfrac{8\pi\sqrt{\rho_{0}}}{9c_{1}},\\
r_{\pm}=&-\dfrac{\mathcal{A}^{\frac{1}{3}}}{36c_{1}}-\dfrac{64\pi^2\rho_{0}}{9c_{1}\mathcal{A}^{\frac{1}{3}}}+\dfrac{8\pi\sqrt{\rho_{0}}}{9c_{1}} \nonumber \\
&\pm\dfrac{i\sqrt{3}}{2}\left(\dfrac{\mathcal{A}^{\frac{1}{3}}}{18c_{1}}-\dfrac{128\pi^2\rho_{0}}{9c_{1}\mathcal{A}^{\frac{1}{3}}}\right),
\end{align}
where
\begin{equation}
\mathcal{A}=4096\pi^3\rho_{0}^{\frac{3}{2}}-\dfrac{2916c_{1}^{2}}{\sqrt{c_{1}}}+108c_{1}\sqrt{\dfrac{729c_{1}^{2}}{\rho_{0}}-2048\pi^{3}}.
\end{equation}
Depending on the value of the parameters, they can be real or imaginary. 
The extremal roots, i.e., where they degenerate to one root, are obtained as
\begin{align}
&c_{1,ext}=9.333\rho_{0},\\
&M_{ext}=\dfrac{0.44881}{\sqrt{\rho_{0}}},\;\;\;\;\to\;\;\;\;
r_{ext}=\left(\dfrac{2}{
c_{1}\sqrt{\rho_{0}}}\right)^{\frac{1}{3}}=\dfrac{0.5984134}{\sqrt{\rho_{0}}}.
\end{align}
Thermodynamically, it corresponds to zero temperature and therefore, zero Hawking radiation which is known as BH remnants for an allowed range of parameters \cite{Chamseddine:2019pux}. We remark that, for $c_{1}>c_{1,ext}$ and $M<M_{ext}$, the horizons disappear. 
 
In Fig. \eqref{wpert1}, we display the behavior of different radii in terms of $c_{1}M^2$. As it can be seen, for  
$0<c_{1}M^2<1.858$, the radius of the object, $R$, is larger than the outer root of the metric function, $r_{+}$.

This means that, before a certain amount of $c_{1}M^2=1.858$, we have a compact relativistic object, and beyond that, we have a BH solution. Indeed, we have $2M(r_{+})/r_{+}=1$  for BHs and for $r>r_{+}$, we have $2M(R)/R<1$ \cite{Liu:2021fit}. We have also shown this behavior as $4\pi M_{in}/C$ (where $C$ is the circumference of the smallest ring that can engulf the object and $M_{in}$ is the mass within the engulfing sphere.) in terms of $c_{1}M^2$. 

For $1.858<c_{1}M^2<2.018$, we have an object with two horizons with $1.367M<r_{\pm}<1.613M$ and $R>r_{\pm}$. In this case, we have a compact star with two horizons inside the object. The transition from star to BH happens at $r_{cr}=1.613M,c_{1cr}M^2=2.018$. For $c_{1}<c_{1ext}$, we have a solution with $r_{\pm}=0$ and $R\neq 0$, which is known as regular ultra-compact star \cite{Volkel:2017kfj}. 

In Fig. \eqref{wpert1}, we also have shown the behavior of the photon sphere radius (gray solid line) versus $c_{1}M^2$. As it can be seen, for $1.522<c_{1}M^2<1.858$, the photon sphere radius does not vanish and there is a lower bound $c_{1, low}$ for which the photon sphere radius is in the interval $r_{ext}\leq r_{ph}\leq r^{sch}_{ph}$. We denote $r^{sch}_{ph}$ as photon sphere radius of the Schwarzschild black hole.

It is interesting to notice that the behavior of $r_{i}/M$ versus $\rho_{0}M^2$ looks the same, as in Fig. \eqref{wpert1}, since $c_{1}$ and $\rho_{0}$ hold the same dimension.

\subsection{Investigating the first solution properties}

In the following, we study the metric in \eqref{hywardlike}, in the BH region {i.e., $R<r_+$.} By expanding the metric function $f$ near the origin, one gets
\begin{equation}
f(r)\sim 1-\dfrac{8\pi\rho_{0}}{3}r^{2}+\mathcal{O}(r^{5}).
\end{equation}
Accordingly, one can interpret the coefficient of $r^{2}$ in the form of a positive effective cosmological constant
$\Lambda_{eff}=8\pi\rho_{0}$, indicating that the solution near the origin behaves as a de Sitter vacuum. By calculating two curvature invariants the Ricci and the Kretschmann scalars near the origin, we obtain
\begin{align}
\lim_{r\to 0}\mathcal{R}=&32\pi \rho_{0}=4\Lambda_{eff},\\
\lim_{r\to 0}\mathcal{K}=&\lim_{r\to 0}\mathcal{R}_{a b c d}\mathcal{R}^{a b c d}=\dfrac{512\pi^2\rho_{0}^2}{3}=\dfrac{8}{3}\Lambda_{eff}^{2}.
\end{align}
One can see that, they are finite at $r=0$, meaning that the first BH solution is free of curvature singularities at the center.  However, the curvature invariant does not have a real physical meaning and only provides a tool for a simple and preliminary analysis of spacetime regularity. In physics, we have to look at the geodesic equations and see whether and where they are well defined. Therefore, a basic requirement for regular spacetime should be the completion of all causal geodesics. It is easy to show that a probe particle can reach the point $r = 0$ in a finite amount of proper time. By extension of the geodesics to the negative values, we find that the point $r = \left(\frac{-1}{c_{1}\sqrt{\rho_{0}}}\right)^{\frac{1}{3}}$ to be the singularity of metric \eqref{hywardlike}. Therefore, this BH is geodesically incomplete and therefore is not an RBH \cite{Zhou:2022yio}.

In the following, we look for a sign of repulsive gravity. It is known that violation of the strong energy conditions is a signature of repulsive gravity. So, the strong energy condition $\rho+P_{r}+P_{t}>0$, is violated at
\begin{equation}
r_{sec}=\left(\dfrac{1}{2c_{1}\sqrt{\rho_{0}}}\right)^{\frac{1}{3}}.
\end{equation}
Similarly, the changes in the sign of the Ricci scalar are a signature of topology change of spacetime. Therefore, a computation shows that the Ricci scalar changes its sign at the radius
\begin{equation}
\mathcal{R}=0\;\;\;\;\to \;\;\;\;r_{R}=\left(\dfrac{2}{c_{1}
\sqrt{\rho_{0}}}\right)^{\frac{1}{3}}.
\end{equation}
The eigenvalues of the curvature $\lambda_{i}$ according to \cite{Luongo:2023aib,Luongo:2014qoa,Luongo:2023xaw,adj1,adj2}, for this metric yields (see Appendix \ref{app:A} for details)
\begin{align}\label{lambdarep}
\lambda_{1}=&-\dfrac{8\pi\rho_{0}(1+c_{1}^2\rho_{0}r^{6}-7c_{1}\sqrt{\rho_{0}}r^3)}{3(1+c_{1}\sqrt{\rho_{0}}r^3)^3},\nonumber\\
\lambda_{4}=&\dfrac{8\pi\rho_{0}}{3(1+c_{1}r^{3}\sqrt{\rho_{0}})}, \nonumber\\
\lambda_{2}=&\lambda_{3}=-\lambda_{5}=-\lambda_{6}=\dfrac{4\pi\rho_{0}(-2+c_{1}r^{3}\sqrt{\rho_{0}})}{3(1+c_{1}r^{3}\sqrt{\rho_{0}})^{2}}.
\end{align}
From \eqref{lambdarep}, we see that $\lambda_{1}$ and $\lambda_{2}$ change their sign when approaching $r=0$, indicating the places at which repulsive gravity becomes dominant. The first extremum that is reached when approaching from infinity is located at $r_{rep}=\left(\frac{8+3\sqrt{6}}{2c_{1}\sqrt{\rho_{0}}}\right)^{\frac{1}{3}}$, which
corresponds to a local minimum of $\lambda_{1}$. The largest extremum belongs to $\lambda_{1}$ and determines the repulsion radius as $r_{rep}$, corresponding
to the onset of repulsive gravity. Moreover, when approaching the source from infinity, the first zero in $\lambda_{1}$ is at $r_{dom}=\left(\frac{7+3\sqrt{5}}{2c_{1}\sqrt{\rho_{0}}}\right)^{\frac{1}{3}}$, which is the point at which repulsive gravity becomes dominant. 

\begin{figure}
\centering
\subfigure{\includegraphics[width=0.98\columnwidth]{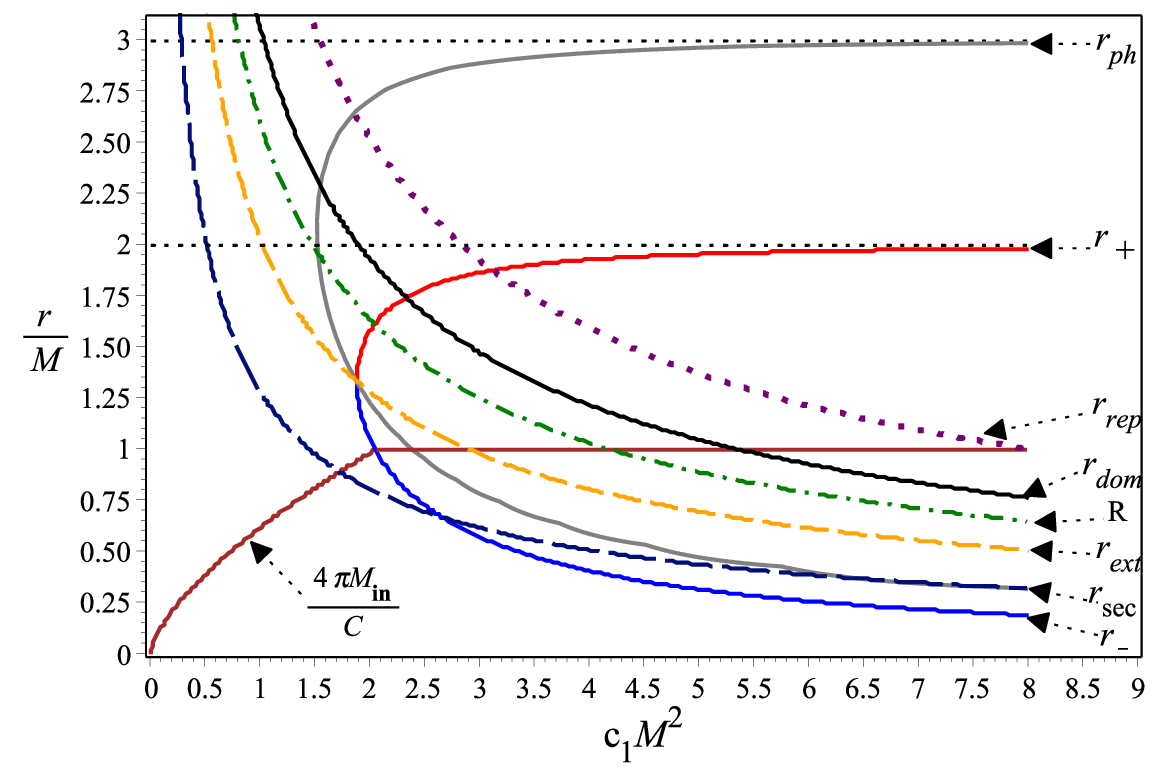}}
\caption{The radial coordinates of key features in the metric \eqref{hywardlike} are depicted here: The behavior of $r_{\pm}/M$, $r_{ext}/M$, $r_{rep}/M$, $r_{ph}/M$ and $R/M$ in terms of $c_{1}M^2$ for the first BH solution.} 
\label{wpert1}
\end{figure}

In Fig.~\eqref{wpert1}, we compare the behavior of different radii. There, a peculiar point is that the $r_{rep}/M$ curve for $3.097<c_{1}M^2$ is smaller than the BH event horizon. For $0<c_{1}M^2<3.097$, the repulsive gravity appears outside the BHs. Hence, one can wonder whether  RBHs and their remnants may be dark energy sources due to their repulsive gravity effects \cite{Calza:2024fzo}. 

Another important aspect is the stability of the Cauchy horizon of these BH solutions. In \cite{Carballo-Rubio:2022kad, Bonanno:2020fgp} by considering the backreaction between the
infalling matter and geometry the stability of Cauchy horizon of black holes has been studied. It has been shown that if the denominator of equation (19) of the paper \cite{Bonanno:2020fgp} is nonlinear, then the Cauchy horizon of the black hole is stable under mass inflation. It is easy to show that the denominator is nonlinear for our metric and therefore the Cauchy horizon of this solution is stable.

\begin{figure*}
\centering
\subfigure{\includegraphics[width=0.95\columnwidth]{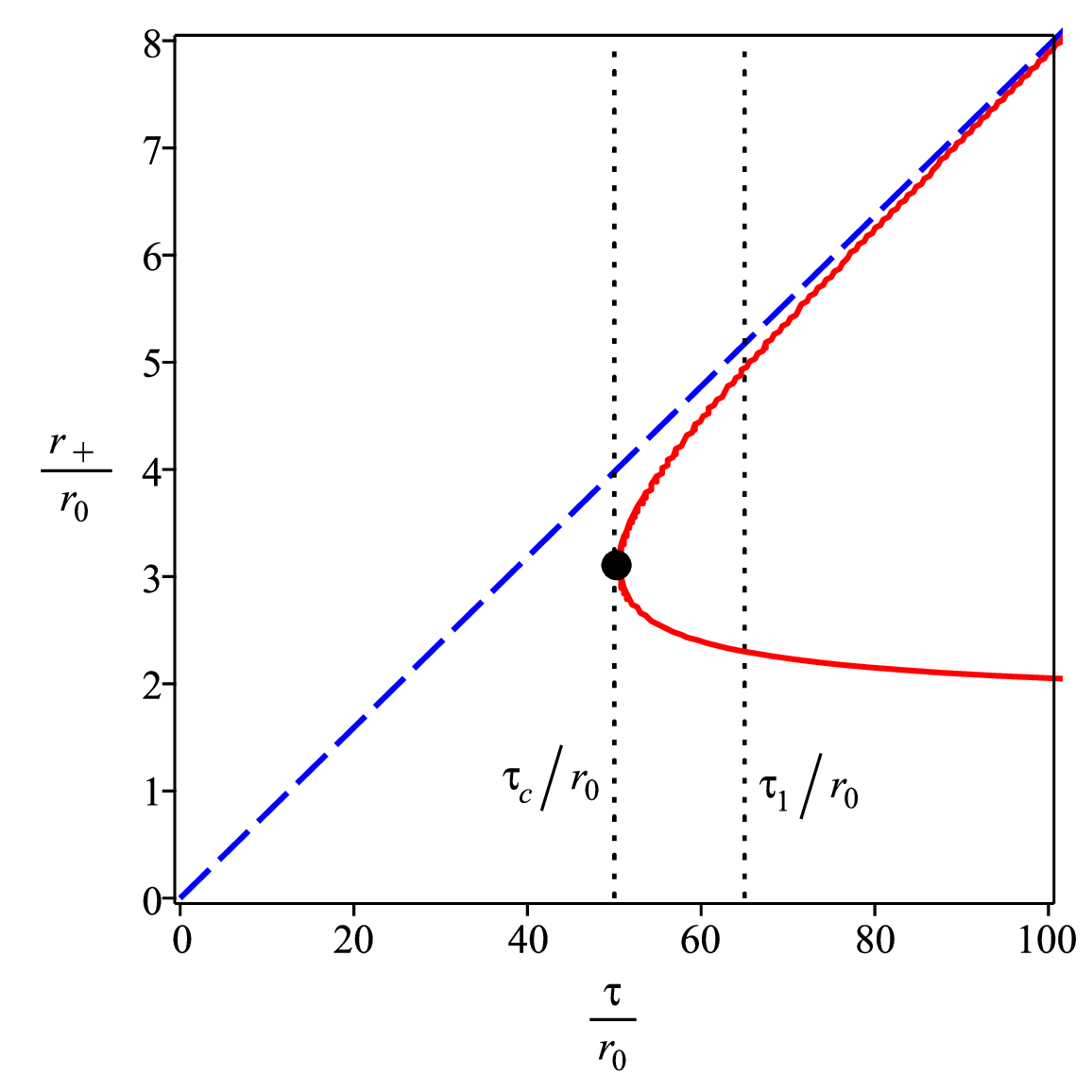}}
\includegraphics[width=0.95\columnwidth]{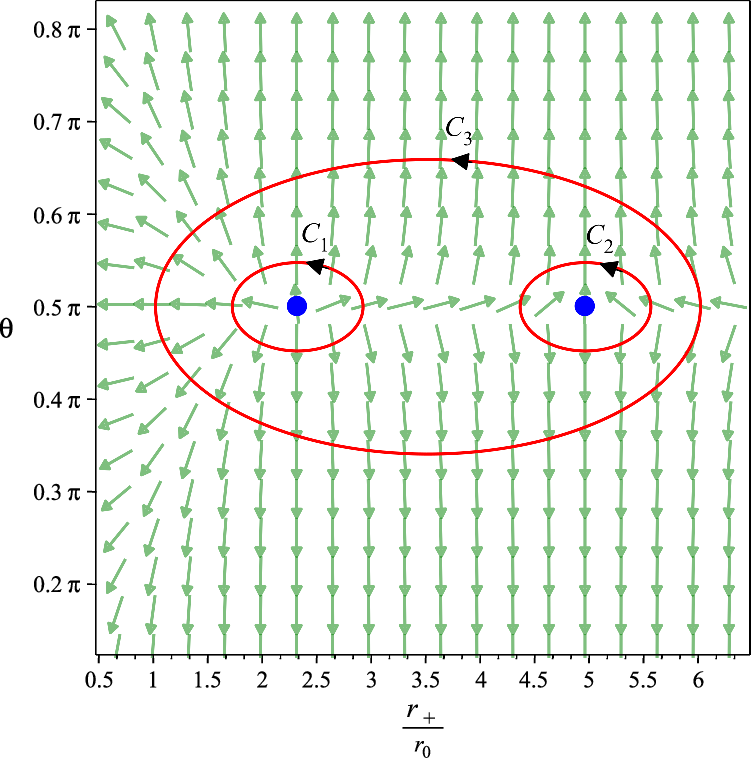}
\caption{Left: Zero points of the vector $\phi$ shown in the $r_{+}-\tau$ for $c_{1}=1$. The blue dashed line is for the Schwarzschild BH and the red solid line is for the first solution. The annihilation point for this BH is represented by the black dot with $\tau_{c}$. There are two RBHs when $\tau=\tau_{1}$. Right: The red arrows represent the unit vector  field $n$ on a the $r_{+}-\theta$ plane for $\tau/r_{0}=65$. The zero points are shown with blue dots and the red contours $C_{i}$ are closed loops surrounding the zero points.} 
\label{zeropoint0}
\end{figure*}

Here, we investigate the thermodynamical topology of an RBH. 
As shown in \cite{Wei:2022dzw,Wei:2021vdx}, one can introduce the free energy as 
\begin{equation}\label{eqftau}
\mathcal{F}=M-\dfrac{\mathcal{S}}{\mathcal{\tau}}=\dfrac{r_{+}}{4}+\dfrac{\sqrt{9c_{1}^{2}r_{+}^4+96\pi}}{12c_{1}r_{+}}-\dfrac{\pi r_{+}^2}{\tau},
\end{equation}
for a BH with mass $M$, entropy $\mathcal{S}=\pi r_{+}^2$ and $\tau$ can be thought of as the inverse temperature of a cavity enclosing the BH. One can construct a vector field $\phi$ as \cite{Wei:2021vdx}
\begin{equation}\label{eqphis}
\phi =\left(\dfrac{\partial \mathcal{F}}{\partial r_{+}},-\cot(\theta)\csc(\theta)\right),
\end{equation}
in which the two parameters $r_{+}$ and $\theta$ obey $0<r_{+}<\infty$ and $0\leq\theta\leq\pi$. By solving the equation $\phi^{r_{+}}=0$, one can obtain a curve on $r_{+}-\tau$ plane. For the first solution, one gets
\begin{equation}
\tau=-\dfrac{8\pi r_{+}^{3}c_{1}\sqrt{9c_{1}^{2}r_{+}^{4}+96\pi}}{32\pi -c_{1}r_{+}^{2}\sqrt{9c_{1}^2r_{+}^4+96\pi}-3c_{1}^2r_{+}^{4}}.
\end{equation}
We show the behavior of $\phi^{r_{+}}$ in the $r_{+}-\tau$ plane in the left figure of Fig.~\eqref{zeropoint0}. For large $\tau=\tau_{1}$, there are two intersection points for the BH. The intersection points exactly satisfy the condition $\tau=1/T$ (where $T$ is the Hawking temperature of a BH) and therefore represent the on-shell BHs. The two intersection points can coincide with each other when $\tau=\tau_{c}$ and vanish when $\tau<\tau_{c}$. Therefore, $\tau_{c}$ is an annihilation point which can be found at $\tau_{c}=50.40245034r_{0}/\sqrt{c_{1}}$ with an arbitrary length scale $r_{0}$ set by the size
of a cavity surrounding the black hole. 

Alternatively, the unit vector field $n=(n^{r},n^{\theta})=(\phi^{r_{+}}/\vert\phi\vert,\phi^{\theta}/\vert\phi\vert)$, is plotted for $\tau=65r_{0}$ in Fig.~\eqref{zeropoint0}b, where we find two zero points at $r_{+}=2.324090959r_{0},\theta=\pi/2$ and at $r_{+}=4.965244619r_{0},\theta=\pi/2$. To determine the winding number of the zero point, we need to construct a closed loop around it and count the changes in the direction of the vector. For this aim, we parametrize the closed loop $C_{i}$ using the angle $s$ as follows
\begin{equation}\label{eqparams}
\dfrac{r_{+}}{r_{0}}=a\cos(s)+r_{c},\;\;\;\;\theta=b\sin(s)+\dfrac{\pi}{2}
\end{equation} 
where $0\leq s\leq 2\pi$ and ($r_{c},\pi/2$) is the center of closed loop. To obtain the change in the direction of the vector, the deflection angle is given by
\begin{equation}\label{eqintome}
\Omega(s)=\oint_{C}\epsilon_{ab}n^{a}n^{b}\partial_{s}n^{b}ds.
\end{equation}
Then, the winding number is defined as $Q=\Omega(2\pi)/2\pi$.

\begin{figure}
\centering
\subfigure{\includegraphics[width=0.95\columnwidth]{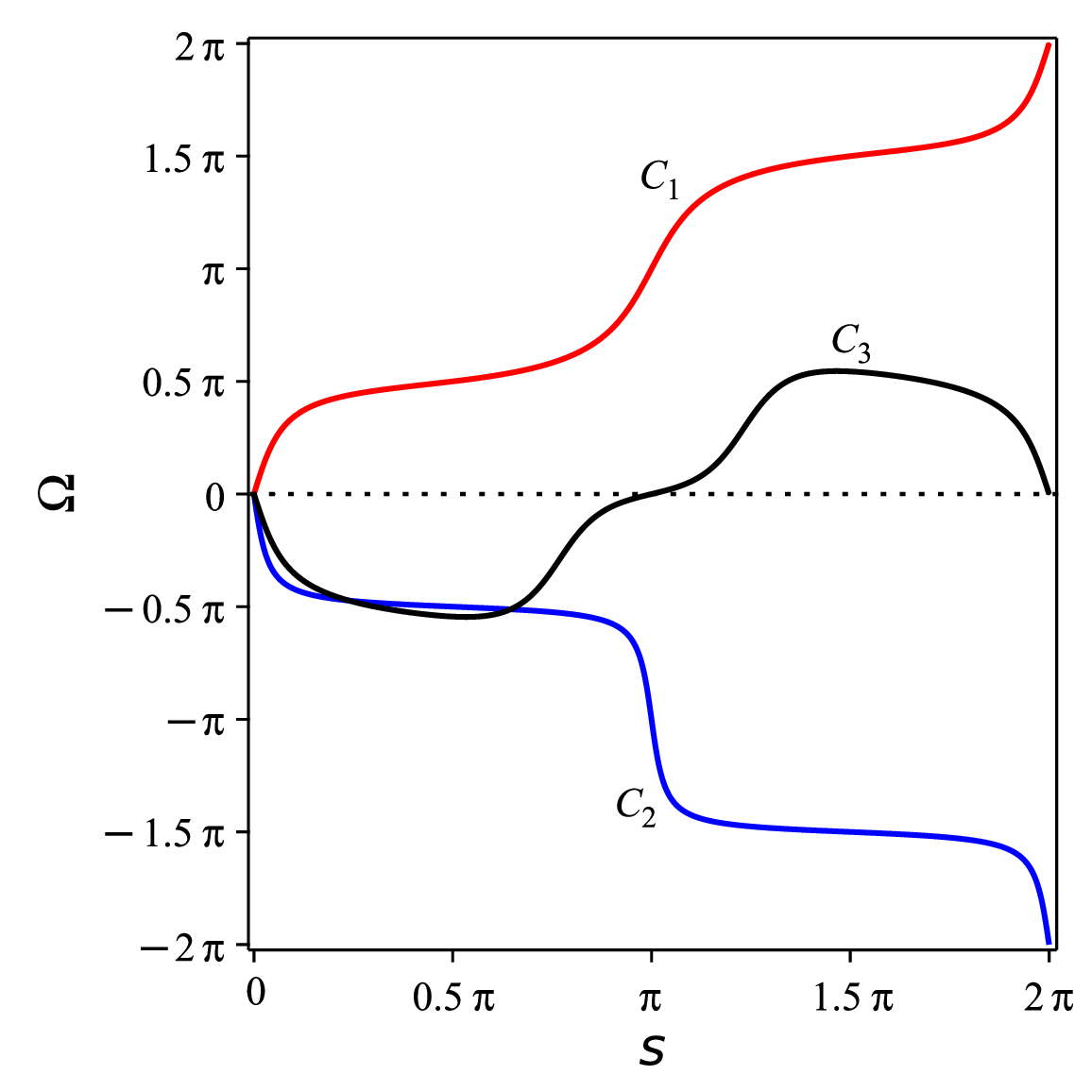}}
\caption{The deflection angle $\Omega$ as a function of $s$ for contours $C_{i}$.} 
\label{defangle}
\end{figure}

Using the method described above, 
we have shown the behavior of deflection angle $\Omega(s)$ for the contours $C_{1}$, $C_{2}$ and $C_{3}$ in Fig~\eqref{defangle}. From the figure, we clearly see that for $C_{1}$ $(C_{2})$ with the increases $s$, $\Omega$ increases (decreases) and approaches $2\pi$ ($-2\pi$) at $s=2\pi$. Thus, we get $Q_{1}=1$  for $C_{1}$ and $Q_{2}=-1$ for $C_{2}$ as expected. Here we have used $a=0.6,b=0.15$.
The contour $C_{3}$ surrounds the two zero points leading to the winding number $Q_{3}=0$. This is because, as can be seen in Fig~\eqref{defangle}, $\Omega$ firstly decreases, then increases, and finally vanishes at $s=2\pi$. Here we have used $a=2.5, b=0.5$.

Finally, using \eqref{eqmassHyward} for the ADM mass and effective cosmological constant $\Lambda_{eff}=3/L^2=8\pi\rho_{0}$, it is easy to rewrite the metric \eqref{hywardlike} as follows:
\begin{equation}
f(r)=1-\dfrac{2Mr^2}{r^3+2ML^2}
\end{equation}
which is the Hayward metric considered in \cite{Hayward:2005gi}. It should be noted that by choosing $m=3/2$ and $\omega_{1}=1$ the Dymnikova RBH solution is also obtained \cite{Dymnikova:2004zc}. Therefore, in this section, we have shown that the Hayward BH can be a solution of Einstein's gravity with an anisotropic perfect fluid with a multi-polytropic equation of state and has a structure like a compact star. But, the remnant of this solution is a regular object with repulsive gravity.

\section{The second 
solution}\label{subsec2.3}
Here in this section, we consider the second solution which can be obtained by choosing 
\begin{subequations}
    \begin{align}
&m=\frac32,\\
&\omega_{1}=3.        
    \end{align}
\end{subequations}
Again, the above choice refers to the need for constant BH mass. Particularly, the tangential pressure, $P_{t}$, in this case, reads
\begin{equation}\label{regular}
P_{t}=3\rho -\dfrac{4\rho^{\frac{3}{2}}}{\sqrt{\rho_{0}}} >0\;\;\to\;\;\begin{cases}
0<\rho<\dfrac{9}{16}\rho_{0}\\
\\
\left(\dfrac{1}{3c_{1}\sqrt{\rho_{0}}}\right)^{\frac{1}{4}}<r<\infty
\end{cases}
\end{equation}

\begin{figure*}
\centering
\includegraphics[width=0.95\columnwidth]{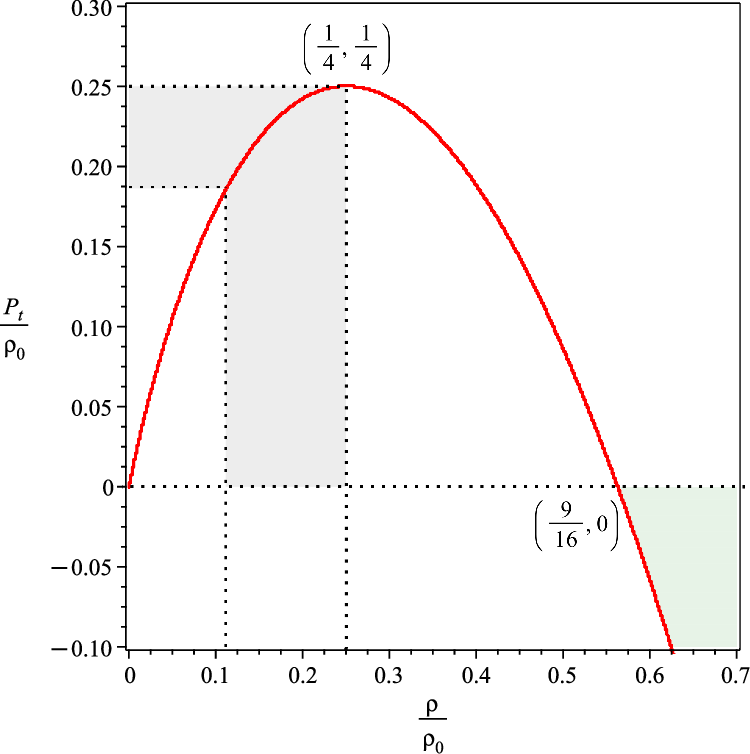}
\includegraphics[width=0.95\columnwidth]{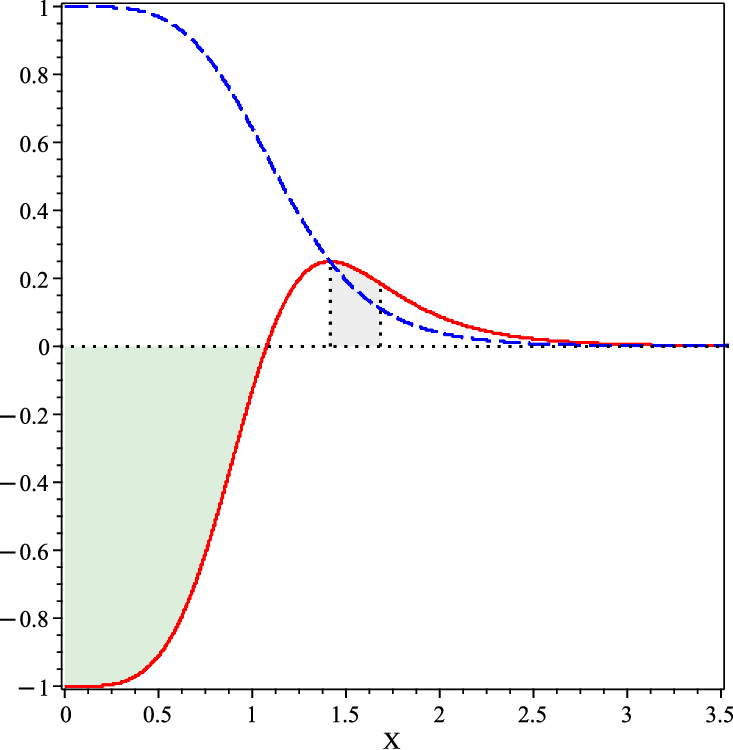}
\caption{Left: The behavior of $P_{t}/\rho_{0}$ in terms of $\rho/\rho_{0}$ for $m=3/2$ and $\omega_{1}=3$. Right: The behavior of $P_{t}/\rho_{0}$ (solid line) and $\rho/\rho_{0}$ (dash line) in terms of dimensionless quantity $X=(\pi^2\rho_{0}M^2)^{\frac{1}{3}}r/M$. The gray shaded regions are the allowed regions for which $1/9\leq \rho/\rho_{0}\leq 1/4$ (left) and $\sqrt{2}\leq X\leq 2^{\frac{3}{4}}$ (right). In the green-shaded regions, SEC violated. } 
\label{wpertfor0}
\end{figure*}

The behavior of $P_{t}/\rho_{0}$ in terms of $\rho/\rho_{0}$ has been shown in Fig.~\eqref{wpertfor0}. For $0<\rho<9\rho_{0}/16$, the SEC condition is satisfied, and for {$\rho=\rho_{0}/4$} or $r=(1/c_{1}\sqrt{\rho_{0}})^{\frac{1}{4}}$, the tangential pressure has a maximum value as {$P_{t,max}=\rho_{0}/4$}. For $\rho>9\rho_0/16$ or $0<r<\left(1/{(3c_{1}\sqrt{\rho_{0}})}\right)^{\frac{1}{4}}$ which is a region close to the center, the SEC is violated. However, the DEC is satisfied when $0\leq \rho\leq \rho_{0}/4$.  Moreover, for $\rho_0/9\leq \rho\leq \rho_0/4$, the speed of sound is $0\leq v_{s}\leq 1$.

To satisfy all the energy conditions (SEC and DEC) and sound speed limit, we find that the energy density must be restricted to $\rho_{0}/9\leq \rho\leq \rho_{0}/4$ which corresponds to $(1/c_{1}\sqrt{\rho_{0}})^{1/4}\leq r\leq (2/(c_{1}\sqrt{\rho_{0}}))^{1/4}$. 
This region is highlighted as a gray area in Fig.~\eqref{wpertfor0}. We also depict $P_t/\rho_0$ and $\rho/\rho_0$ as function of $X=(\pi\rho_{0}M^2)^{\frac{1}{3}}r/M$ in the right panel of Fig.~\eqref{wpertfor0}. We find that the energy density decreases as a function of $r$. In contrast, the tangential pressure reaches its maximum at $X=\sqrt{2}$ and then vanishes at large $r$.

\begin{figure}
\centering
\subfigure{\includegraphics[width=0.98\columnwidth]{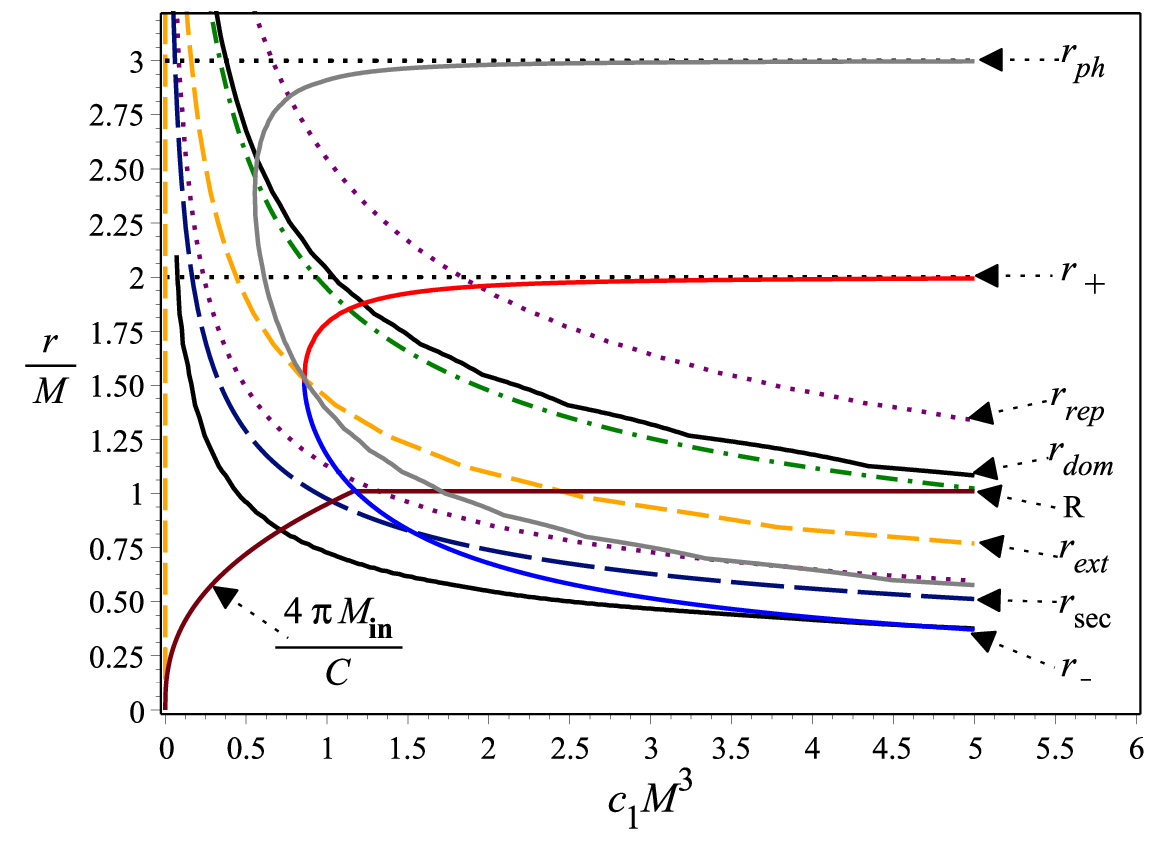}}
\caption{The radial coordinates of key features in the metric function \eqref{eqsol4} are depicted here: The behavior of $r_{\pm}/M$, $r_{ext}/M$, $r_{rep}/M$, $r_{ph}/M$ and $R/M$ in terms of $c_{1}M^3$ for the second BH solution.} 
\label{wpert4}
\end{figure}

In this case, the metric function is obtained as
\begin{widetext}
\begin{align}\label{eqsol4}
f(r)=1-\dfrac{2\pi\rho_{0}r^2}{1+c_{1}\sqrt{\rho_{0}}r^4}+\dfrac{\pi\sqrt{2\rho_{0}}}{2c_{1}\mathcal{E}^{\frac{1}{4}}r}\Bigg[\ln{\left(\dfrac{r^2+\sqrt{2}\mathcal{E}^{\frac{1}{4}}r+\sqrt{\mathcal{E}}}{r^2-\sqrt{2}\mathcal{E}^{\frac{1}{4}}r+\sqrt{\mathcal{E}}}\right)^{\frac{1}{2}}} -\arctan{\left(\dfrac{\sqrt{2}r}{\mathcal{E}^{1/4}-\mathcal{E}^{-1/4} r^2}\right)}\Bigg],
\end{align}
\end{widetext}
here $\mathcal{E}=(c_{1}\sqrt{\rho_{0}})^{-1}$. 
The asymptotic behavior of the metric function is given
\begin{align}
f(r) &\sim  1-\dfrac{\pi^2\sqrt{2}\rho_{0}^{\frac{5}{8}}}{2c_{1}^{\frac{3}{4}}r}+\mathcal{O}(r^{-3}),
\end{align}
here, it is easy to show that the dimension of $c_{1}$ is $[c_{1}]=L^{-3}$. The total mass of the system is given by
\begin{equation}
M=4\pi\int_{0}^{\infty}r^2\rho(r)dr=\dfrac{\sqrt{2}\pi^2\rho_{0}^{\frac{5}{8}}}{4c_{1}^{\frac{3}{4}}}.
\end{equation}
Accordingly, the effective radius ($R$) of the object in the large radius can be approximated as
\begin{equation}
R\sim \left(\dfrac{M}{\rho_{0}}\right)^{\frac{1}{3}}=\left(\dfrac{\sqrt{2}\pi^2}{4c_{1}^{\frac{3}{4}}\rho_{0}^{\frac{3}{8}}}\right)^{\frac{1}{3}}.
\end{equation}
The metric \eqref{eqsol4} is a BH solution if the largest root of the metric ($r_{+}$) is greater than the above radius ($R$).
In Fig.~\eqref{wpert4}, the behavior of different properties of the solution in terms of dimensionless quantity $c_{1}M^3$ has been shown. For $0\leq c_{1}M^3\leq 1.123$, the radius of the object ($R$) is larger than the larger root of the metric ($r_+$) \eqref{eqsol4}. and we have also shown this in another language as $4\pi M_{in}/C<1$ (brown solid curve). This means, in the mentioned range, we have an ultra-compact star. For $c_{1}M^3>1.123$, since $r_{+}>R$, therefore we have a BH solution with $4\pi M_{in}/C=1$. In the figure, we also have shown the radius of repulsive gravity $r_{rep}$ (dotted line) and dominant repulsive gravity ($r_{dom}$ black solid line).

As can be seen, for $0\leq c_{1}M^3\leq 1.933$ the radius of repulsive gravity is larger than the radius of an event horizon. Similar to the first solution, this means that the remnant of the BH has a repulsive gravity. In Fig.~\eqref{wpert4}, the behavior of photon sphere radius has been shown in a solid gray line.  For the range $0.522\leq c_{1}M^3\leq 0.849$, the radius of the photon sphere is not zero $r_{ext}\leq r_{ph}\leq r_{ph}^{sch}$. It should be noted that most of the results in Fig.~\eqref{wpert4} are obtained numerically. This is because the solution \eqref{eqsol4} is complicated and finding associated radii cannot be done analytically. {
Before analysing the near origin behaviour, let us consider the inner horizon of the BH. To check the stability of the Cauchy horizon of the solution under the passing of matter, we shall consider the denominator of the relation (19) of the paper \cite{Bonanno:2020fgp}. Since the denominator is nonlinear for our metric, the solution is stable under mass inflation.}

\begin{figure}
\centering
\subfigure{\includegraphics[width=0.98\columnwidth]{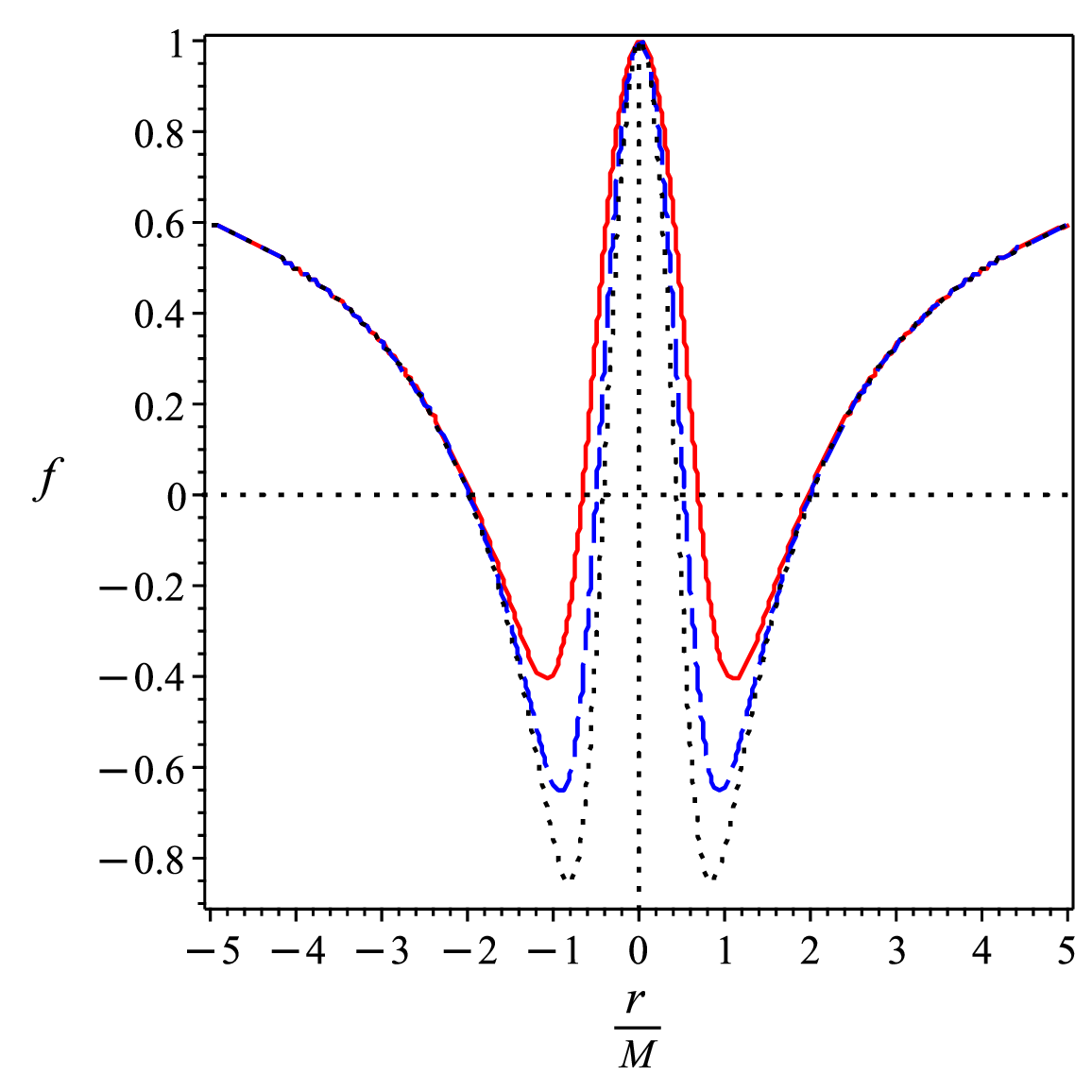}}
\caption{The behavior of metric function \eqref{eqsol4} in terms of $r/M$, for $c_{1}M^3=\textcolor{red}{2},\textcolor{blue}{3},4$.} 
\label{frplott}
\end{figure}

The behavior of metric near the origin is given by 
\begin{equation}
f(r)\sim  1-\dfrac{8\pi\rho_{0}}{3}r^{2}+\mathcal{O}(r^3).
\end{equation}
Here, one can interpret the coefficient of $r^2$ as an effective cosmological constant as $\Lambda_{eff}=8\pi\rho_{0}$.
The behavior of the Ricci and the Kretschmann scalar near the origin is given by
 \begin{align}
\lim_{r\to 0} \mathcal{R}=&32\pi\rho_{0}=4\Lambda_{eff},\nonumber\\
\lim_{r\to 0} \mathcal{K}=&\dfrac{512\pi^2\rho_{0}^2}{3}=\dfrac{8}{3}\Lambda_{eff}^{2}.
 \end{align}
One can see that they are finite at $r = 0$, meaning that the second BH solution is free of
curvature singularities at the center. However, it is easy to show that the spacetime metric \eqref{eqsol4}, is also geodesically complete for both massive and massless particles. Since the metric function \eqref{eqsol4} is continuous across $r=0$ and the effective potential is finite everywhere, we can extend the metric across $r=0$ to negative values of $r$ as well (see Fig.~\ref{frplott}). Thus a test particle can move from $-\infty$ to $+\infty$
in this spacetime, and this takes an infinite amount of proper time. Therefore, this BH is, indeed, regular BH \cite{Zhou:2022yio}.

\subsection{Properties of the second solution}

\begin{figure*}
\centering
\subfigure{\includegraphics[width=0.95\columnwidth]{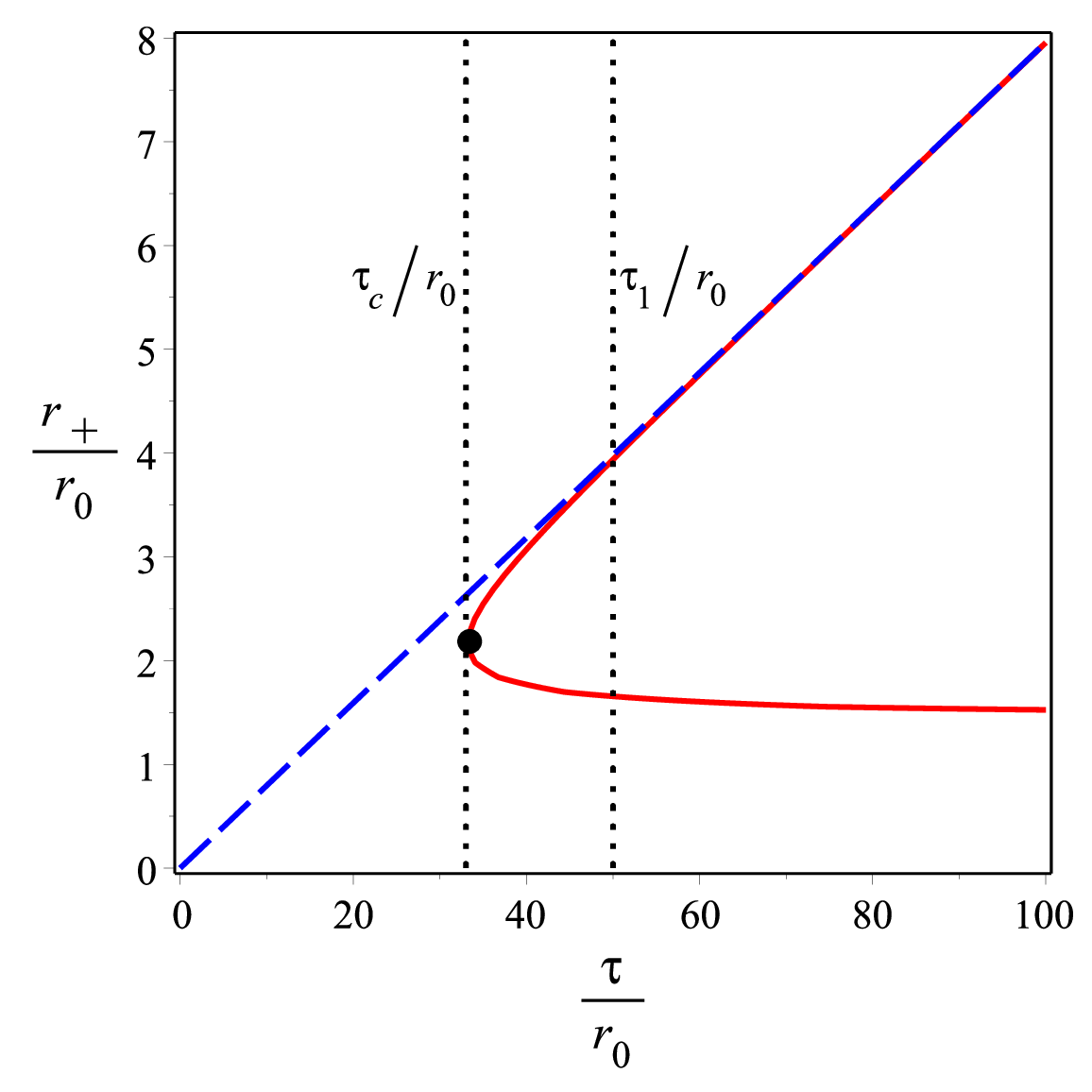}}
\includegraphics[width=0.95\columnwidth]{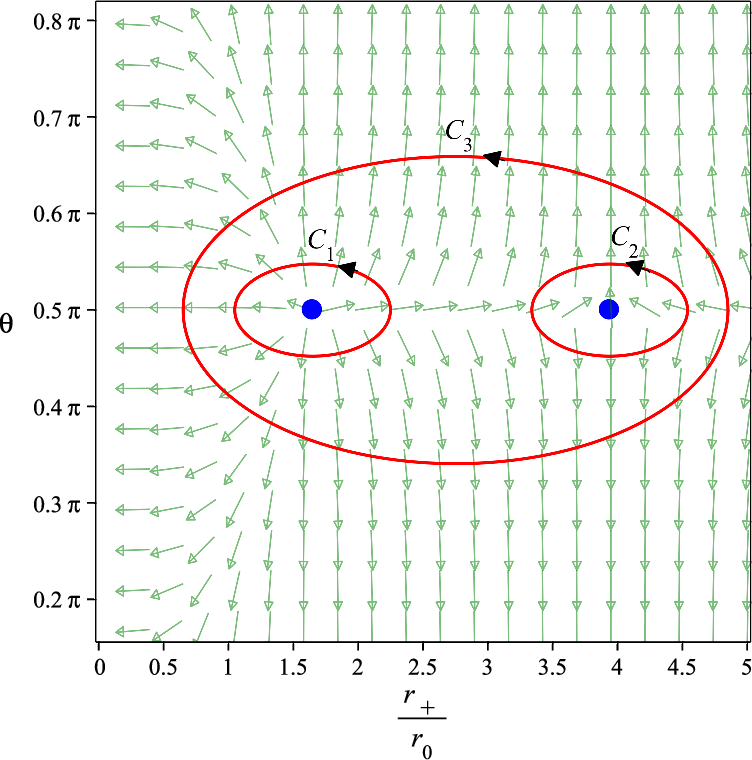}
\caption{Left: Zero points of the vector $\phi$ shown in the $r_{+}-\tau$ for $c_{1}=1$. The blue dashed line is for the Schwarzschild BH and the red solid line is for the second solution. The annihilation point for this BH is represented by the black dot with $\tau_{c}=33.022$. There are two RBHs when $\tau=\tau_{1}$. Right: The red arrows represent the unit vector  field $n$ on a the $r_{+}-\theta$ plane for $\tau/r_{0}=50$. The zero points are shown with blue dots and the red contours $C_{i}$ are closed loops surrounding the zero points. } 
\label{zeropertfor0}
\end{figure*}

\begin{figure}
\centering
\subfigure{\includegraphics[width=0.95\columnwidth]{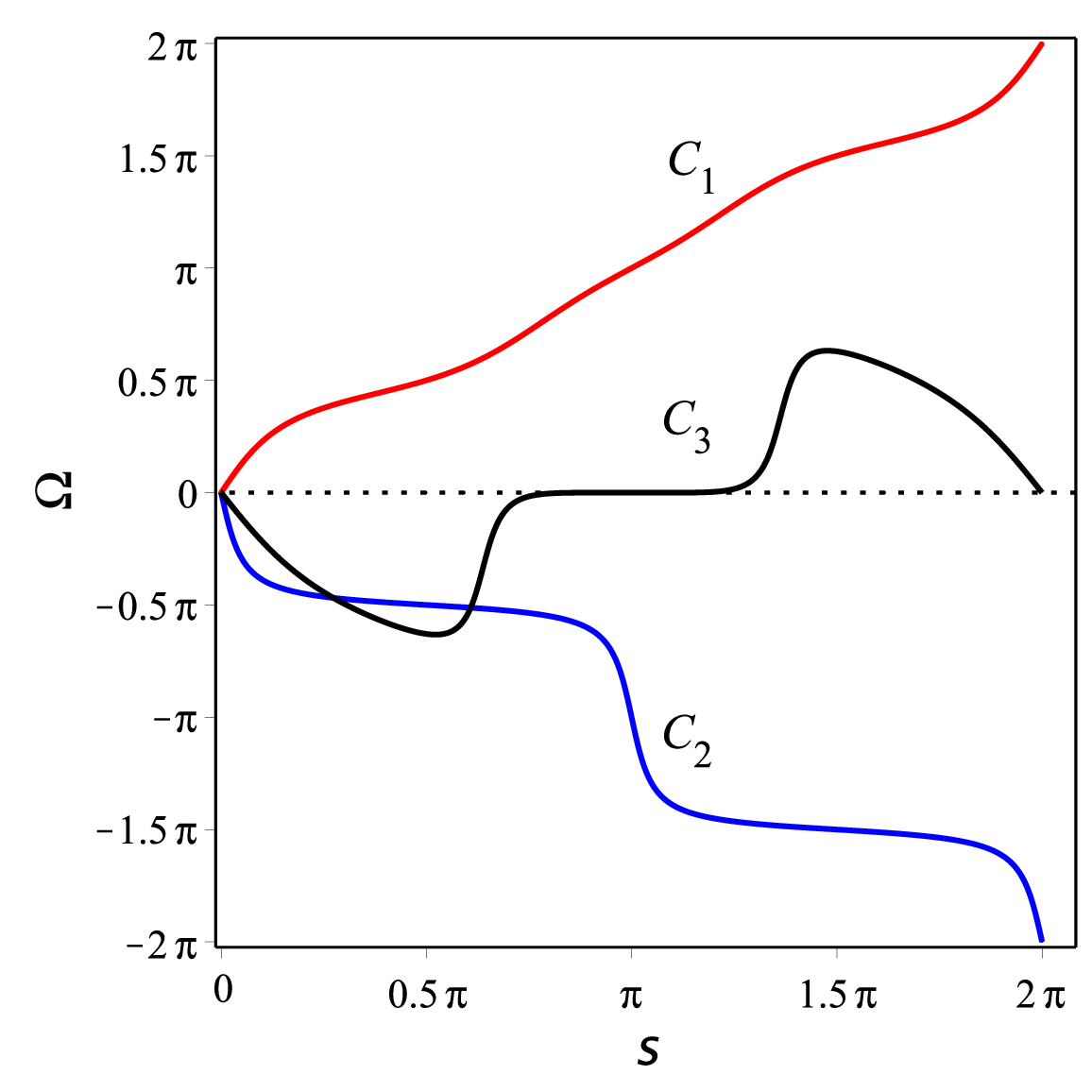}}
\caption{The deflection angle $\Omega$ as a function of $s$ for contours $C_{i}$.} 
\label{defanglesec}
\end{figure}

In analogy to the previous section, here we study the thermodynamic topology of RBH. Based on \eqref{eqftau} and \eqref{eqphis}, we can obtain the free energy and components of the vector field $\phi$. Thus, one can plot the corresponding $r_{+}-\tau$ diagram, the vector field $\phi$, and the deflection angle which are shown in Figs.  \eqref{zeropertfor0} and \eqref{defanglesec}.

For large $\tau=\tau_{1}$, there are two intersection points for the BH. The two intersection points coincide with each other when $\tau=\tau_{c}$ and then vanish when $\tau<\tau_{c}$. The unit vector field $n$, is plotted for $\tau=50r_{0}$ in Fig.~\eqref{zeropoint0}b. We find two zero points at $r_{+}=1.648r_{0},\theta=\pi/2$ and at $r_{+}=3.938 r_{0},\theta=\pi/2$. The winding number of zero points can be determined using \eqref{eqintome}, after parameterizing the contours using \eqref{eqparams}.
The behavior of deflection angle $\Omega(s)$ for the contours $C_{1}$, $C_{2}$ and $C_{3}$ has been shown in Fig.~\eqref{defanglesec}. From the figure, we clearly see that for $C_{1}$$(C_{2})$ with the increases $s$, $\Omega$ increases (decreases) and approach $2\pi$$(-2\pi)$ at $s=2\pi$. Thus, we get $Q_{1}=1$  for $C_{1}$ and $Q_{2}=-1$ for $C_{2}$ as expected. Here we have used $a=0.6,b=0.15$.
The contour $C_{3}$ surrounds the two zero points leading to the winding number $Q_{3}=0$. Because, as can be seen in Fig.~\eqref{defanglesec}, by increasing $s$, $\Omega$ firstly decreases and then increases with $s$. Finally, it vanishes at $s=2\pi$. Here we have used $a=2.1, b=0.5$. As observed, the topological number of this RBH is the same as the previous one, 0, indicating that both belong to the same topological class.

Then, we study the dynamical stability of the BH using the perturbation of the space-time utilizing a real massless scalar field. The evolution of a scalar field in curved spacetime is given by Klein-Gordon equation i.e., $\nabla_{\mu}\nabla^{\mu}\Phi=0$. In a static spherically symmetric background, the scalar field can be assumed to take the following form
\begin{align}
    \Phi &= \frac{R(r)}{r}e^{i\omega t}Y(\theta,\phi),
\end{align}
where $Y(\theta,\phi)$ is spherical harmonics. The Klein-Gordon equation reads
\begin{align}
    \frac{d^2R}{dr_{\ast}^2} + \left[\omega^2-f\left(\frac{l(l+1)}{r^2} + \frac{f'}{r}\right)\right]R &=0,
\end{align}
where $l$ is orbital quantum number and we have introduced tortoise coordinate $dr_{\ast}=dr/f$. The stability of the BH solution is determined via computing quasinormal modes (QNMs). The corresponding boundary condition is that only ingoing scalar wave is allowed at the event horizon and purely outgoing wave at asymptotic infinity. As a result, we obtain a discrete complex eigenfrequency ($\omega=\omega_r + i\omega_i$) or quasinormal frequency. 

To obtain quasinormal frequency, we follow Mashhoon's method which offers analytic way to determine $\omega$ via the approximation of P\"oschl-Teller potential \cite{vf}. This method is widely used in obtaining quasinormal frequencies of several spherical symmetric BH spacetimes \cite{vf,Cardoso:2003sw,Molina:2003ff,Ponglertsakul:2018smo,Burikham:2020dfi,Wuthicharn:2019olp}. Additionally, with some alteration, Mashhoon's method is also found to be useful when determining the frequency spectrum of slowly rotating black holes \cite{Ferrari:1984ozr,Ponglertsakul:2020ufm}. 

\begin{figure*}
\centering
\subfigure{\includegraphics[width=0.95\columnwidth]{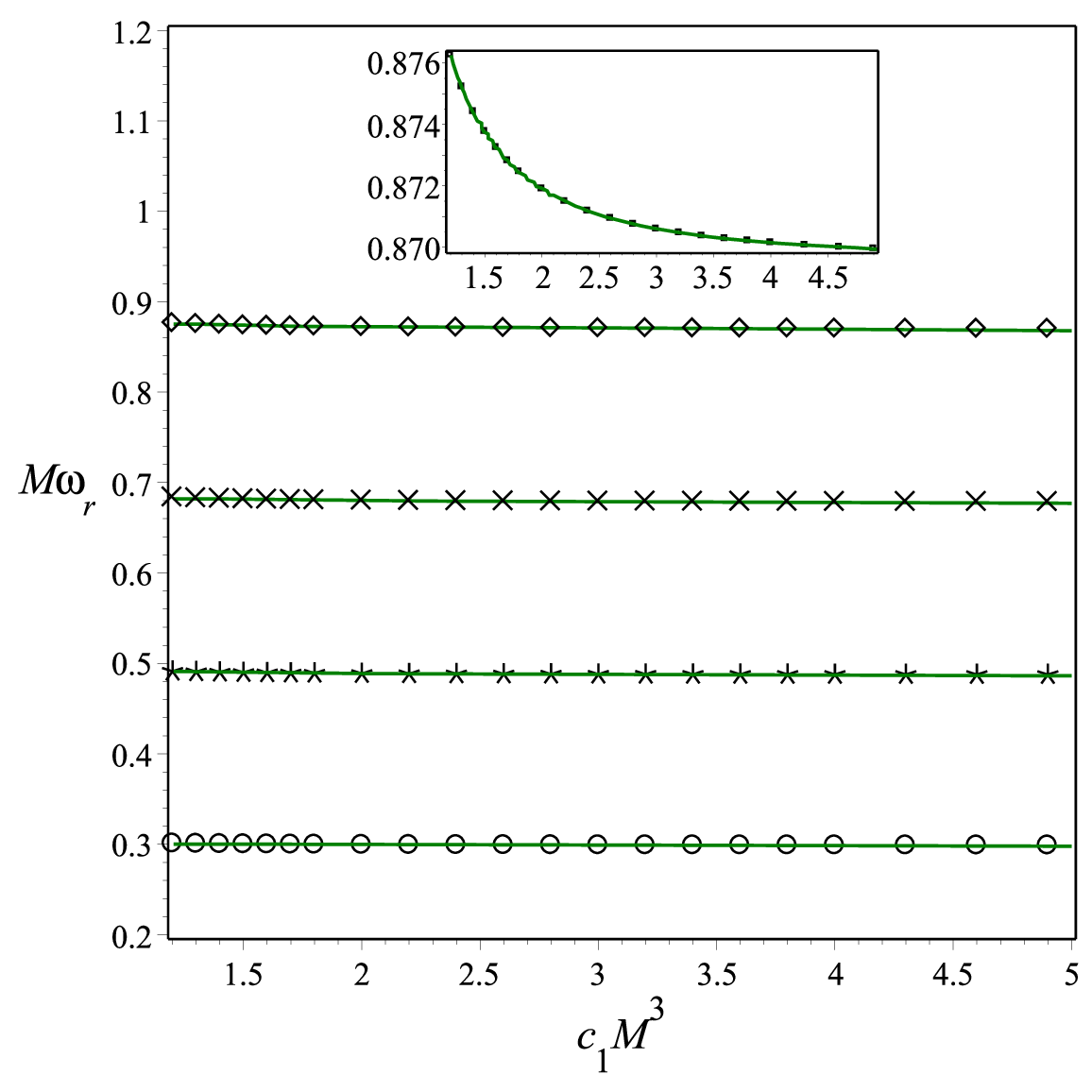}}
\subfigure{\includegraphics[width=0.95\columnwidth]{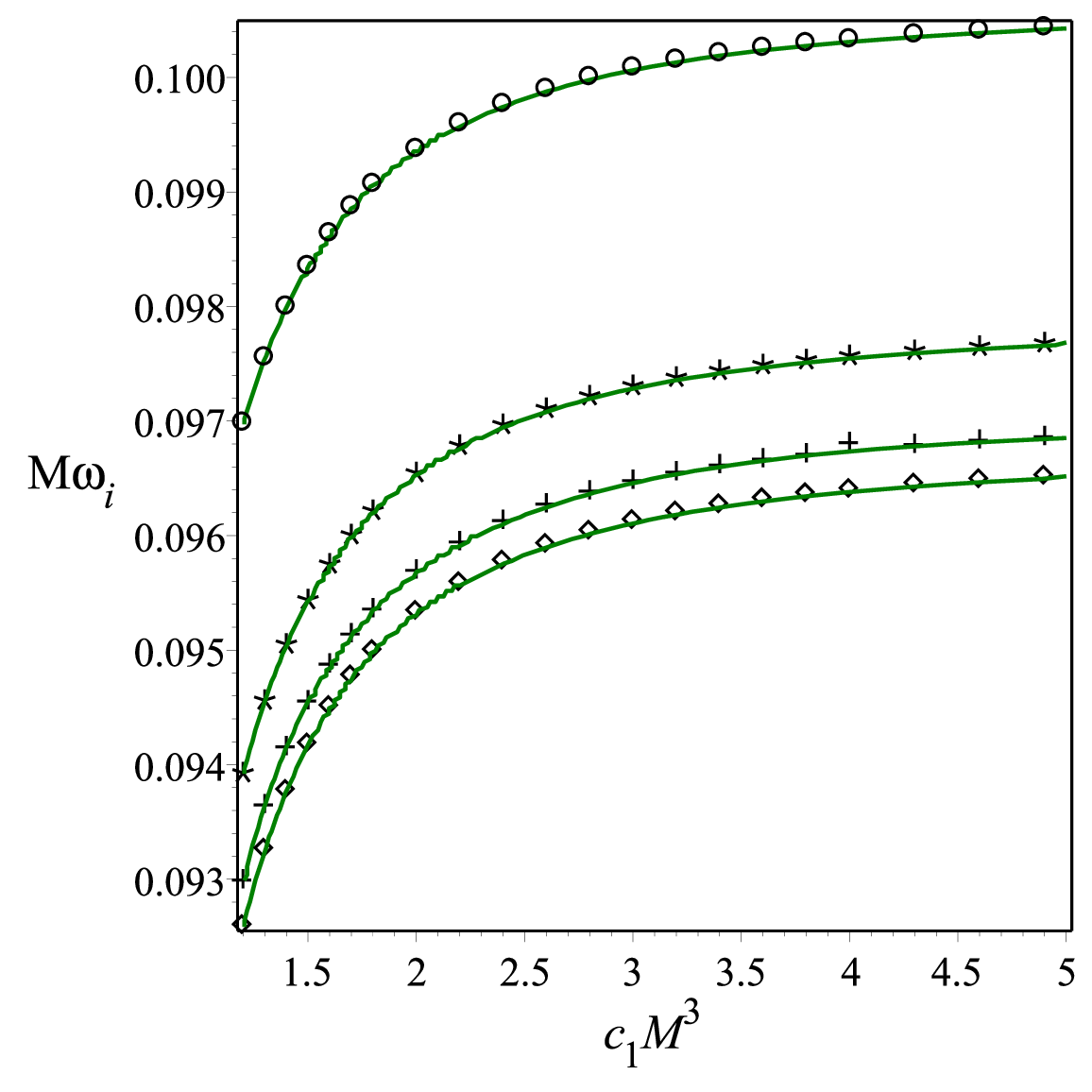}}
\caption{Left: The real part of the quasinormal frequency for $l=1,2,3,4$ (bottom to top). Right: The imaginary part of the quasinormal frequency for $l=1,2,3,4$ (top to bottom).} 
\label{wrc1plot}
\end{figure*}

In Fig.~\ref{wrc1plot}, we have given the results related to calculating quasinormal frequencies where the real and imaginary part of the frequencies are plotted against $c_1M^3$. The first conclusion that can be drawn from this figure is that the BH remains stable under perturbation. This is because the imaginary part of the frequencies is positive rendering exponential decaying of the perturbation. An additional property is that as $l$ increases, the real part of the frequency increases and the imaginary part decreases. This means that the wave is less damped as $l$ increases. Also, as it is clear in the figure, with the increase of $c_{1}$, the real part decreases, and the imaginary part increases. {This means that $c_{1}$ increases the stability of the black hole.}

One final remark is, we find that by choosing $m=\omega_{1}=2$, one can obtain a new RBH solution that has almost the same behavior as the solutions presented in this section and it is possible to study it analytically.

\section{Final outlooks and perspectives}\label{concluding}

In this paper, we construct regular black hole solutions in general relativity by assuming anisotropic fluid as a source. Indeed, RBHs have been usually proposed based on purely-phenomenological grounds, for example, lying on theories of non-linear electrodynamics that can generalize the standard Hilbert-Einstein action. 

In so doing, we provide a classical strategy to obtain RBHs and start from the widely-used TOV equation, imposing polytropic equations of state.

More precisely, we infer an RBH solution by considering anisotropic perfect fluid as a source with a multi-polytropic equation of state. Nevertheless, the structure of the anisotropic pressure consists of radial and transverse components, exhibiting vacuum energy and a multi-polytropic counterpart with a linear term plus a power-law contribution. 

To do so, we solve the system of equations of motion by setting certain numerical values for the model parameters denoted as $m_{1},\omega_{1},c_{1}$, and $\rho_{0}$. These parameters are chosen such that they fulfill physical conditions in order to reproduce viable RBH solutions.

Accordingly, we thus derive two analytical solutions prompting a de Sitter core, i.e., behaving regularly at $r = 0$ and smoothly approaching an asymptotically flat spacetime for large radii. Afterward, we check the regularity of the solutions using curvature scalars and the geodesic completeness of the aforementioned spacetimes. 

Interestingly, we seek regions in which gravity switches its sign, within the frameworks of the underlying metrics. Thus, using the eigenvalues of the Riemann curvature tensor, decomposed into its irreducible representation, we consider the effects of repulsive gravity for these metrics to emphasize the peculiarities of the solutions. Moreover, we study the stability of the RBHs under the flow of energy through the inner horizon and, furthermore, we investigate BH thermodynamical stability by counting winding numbers in the thermodynamic parameter space. 

We find that these solutions belong to the class of solutions with the topological number $W=0$. Thus, we conclude that our RBHs can be described by polytropic models similar to other compact objects due to the absence of singularities, suggesting the use of such spacetimes to model physical objects. 

As a perspective, we notice that these solutions can be further investigated, for instance, regarding the dynamic stability of these solutions against superradiance scattering and thermodynamics of the BH solutions. In addition, the accretion disk regions \cite{accretion1,accretion2,accretion3}, expected from the aforementioned spacetimes, will be explored. Last but not least, future works can focus on finding out new metrics, through the same procedure.

\section*{Acknowledgements}
This research has received funding support from the NSRF via the Program Management Unit for Human Resource and Institutional Development, Research and Innovation grant number $B13F670063$. OL acknowledges financial support from the  Fondazione  ICSC, Spoke 3 Astrophysics and Cosmos Observations. National Recovery and Resilience Plan (Piano Nazionale di Ripresa e Resilienza, PNRR) Project ID CN$\_$00000013 "Italian Research Center on  High-Performance Computing, Big Data and Quantum Computing"  funded by MUR Missione 4 Componente 2 Investimento 1.4: Potenziamento strutture di ricerca e creazione di "campioni nazionali di R$\&$S (M4C2-19 )" - Next Generation EU (NGEU)
GRAB-IT Project, PNRR Cascade Funding
Call, Spoke 3, INAF Italian National Institute for Astrophysics, Project code CN00000013, Project Code (CUP): C53C22000350006, cost center STI442016. 

\appendix

\section{Repulsive gravity}\label{app:A}

In this Appendix, we review the concept of repulsive gravity, used throughout the text and, in particular, its invariant description, in terms of the eigenvalues of the Riemann curvature tensor. 

In general, repulsive gravity is certified by investigating the effective potential of any metric, i.e., as the effective potential changes sign, one can infer regions in which the sign of gravity becomes repulsive. This procedure, albeit physically acceptable, is coordinate-dependent. So, seeking an invariant formalism toward repulsive regions is of utmost importance. 

As a naive example, the simplest repulsive solution is offered by a Schwarzschild metric where the gravitational charge has a reversed sign, namely when one assumes the mass to satisfy $M\rightarrow-M$, see  \cite{Luongo:2014qoa}. The repulsive sign of gravity cannot be reached as the first-order curvature invariant, the Ricci scalar, vanishes and all the higher-order geometric invariants keep hidden the sign of $M$, see \cite{Luongo:2023aib}. 

Accordingly, to obtain the Riemann curvature tensor in a local orthonormal frame, we choose the orthonormal tetrad $e^{a}$ as
\begin{equation}
ds^{2}=\eta_{ab}e^{a}\wedge e^{b},
\end{equation}
with $\eta_{ab}=diag(-1,1,1,1)$ and $e^{a}=e^{a}_{b}dx^{\mu}$. The first and the second Cartan structure equations are
\begin{align}
&de^{a}+\omega^{a}_{b}\wedge e^{b}=0,\nonumber\\
&\Omega^{a}_{b}=d\omega^{a}_{b}+\omega^{a}_{c}\wedge\omega^{c}_{b}=\dfrac{1}{2} R^{a}{}_{bcd}e^{c}\wedge e^{d},
\end{align}
allowing to introduce the connection 1-form $\omega^{a}_{b}$ and the curvature 2-form $\Omega^{a}_{b}$, which determines the tetrad components of the curvature tensor $R^{a}{}_{bcd}$. 

Further, to compute the curvature eigenvalues, it is convenient to introduce the convention \cite{Misnerbook}
\begin{align}
&1\to 01,\;\;\;2\to 02,\;\;\;3\to 03,\;\;\;4\to 23,\nonumber\\
&5\to 31,\;\;\;6\to 12.
\end{align}
Then, the curvature eigenvalues are easily computed as the eigenvalues of the ($6\times 6$)- matrix $R_{AB}$. Considering the spacetime geometry \eqref{metricasl}, after some computations, it is easy to obtain eigenvalues of curvature two form as follows \cite{Luongo:2014qoa}:
\begin{align}
\lambda_{1}=&\dfrac{2hf f^{\prime\prime}+ff^{\prime}h^{\prime}-hf^{\prime 2}}{4f^2},\nonumber\\
\lambda_{2}=&\lambda_{3}=-\dfrac{h^{\prime}}{2r},\;\;\lambda_{4}=\dfrac{1-h}{r^{2}},\nonumber\\
\lambda_{5}=&\lambda_{6}=\dfrac{hf^{\prime}}{2fr}.
\end{align}
These eigenvalues encapsulate all the information regarding curvature exhibit scalar behavior under coordinate transformations and appear coordinate-independent. 

To identify repulsive gravity one adopts the following steps. 

\begin{itemize}
    \item[-] Searching for the sign change of \emph{at least one eigenvalue} that implies the transition from attractive to repulsive gravity. Consequently, the eigenvalue zeros mark the domain where repulsive gravity becomes dominant.
    \item[-] Finding extremal points in \emph{at least} one eigenvalue, indicating the onset of repulsive effects, as determined by the condition:
\begin{equation}
\dfrac{\partial\lambda_{i}}{\partial r}=0.
\end{equation}
\end{itemize} 

The above procedure has been used across the manuscript to identify the repulsive regions and has been widely adopted to cosmological contexts \cite{Luongo:2015zaa} and gravitational solutions \cite{Luongo:2014qoa, Luongo:2023aib, Sajadi:2023ybm, Luongo:2023xaw}, with promising outcomes indicating that the spacetime alone can exhibit regions where gravity changes its sign.

\end{document}